\documentclass{article}
\usepackage{ijcai25}

\usepackage{times}
\usepackage{soul}
\usepackage{url}
\usepackage[hidelinks]{hyperref}
\usepackage[utf8]{inputenc}
\usepackage[small]{caption}
\usepackage{graphicx}
\usepackage{amsmath}
\usepackage{amsthm}
\usepackage{booktabs}
\usepackage{algorithm}
\usepackage{algorithmic}
\usepackage[switch]{lineno}
\usepackage{amssymb}
\usepackage[most]{tcolorbox}
\usepackage{listings}

\newtheorem{example}{Example}

\title{AgentDAO: Synthesis of Proposal Transactions Via Abstract DAO Semantics}

\author{
Lin Ao$^1$
\and
Han Liu$^1$\And
Huafeng Zhang$^1$\\
\affiliations
$^1$Weaver Technology\\
\emails
\{aolin118, liuhan0518, ron.huafeng\}@gmail.com
}

\begin{document}

\maketitle

\begin{abstract}
While the trend of decentralized governance is obvious (cryptocurrencies and blockchains are widely adopted by multiple sovereign countries), initiating governance proposals within Decentralized Autonomous Organizations (DAOs) is still challenging, i.e., it requires providing a low-level transaction payload, therefore posing significant barriers to broad community participation. To address these challenges, we propose a multi-agent system powered by Large Language Models with a novel Label-Centric Retrieval algorithm to automate the translation from natural language inputs into executable proposal transactions. The system incorporates DAOLang, a Domain-Specific Language to simplify the specification of various governance proposals. The key optimization achieved by DAOLang is a semantic-aware abstraction of user input that reliably secures proposal generation with a low level of token demand. A preliminary evaluation on real-world applications reflects the potential of DAOLang in terms of generating complicated types of proposals with existing foundation models, e.g. GPT-4o.

\end{abstract}

\section{Introduction}

Decentralized Autonomous Organizations (DAOs)\cite{wang2019decentralized} are blockchain-based entities governed by encoded rules and protocols, designed to operate without reliance on centralized authorities. DAOs leverage smart contracts—self-executing programs on blockchain networks to facilitate transparent, trustless, and autonomous decision-making processes. In the governance framework of a DAO, proposals \cite{hassan2021decentralized} serve as the primary mechanism through which members initiate, discuss, and enact changes or decisions. A proposal outlines a specific action or modification, such as funding allocation, protocol adjustments, or governance updates, and is submitted for review by the DAO community.

While the trend of decentralized governance is obvious (cryptocurrencies and blockchains are widely adopted by multiple sovereign countries), initiating governance proposals within Decentralized Autonomous Organizations (DAOs) is still challenging. A major challenge in the mass adoption of DAOs is the requirement for expertise in high-level techniques to initiate proposals. Governance proposals must include transaction data detailing steps to interact with smart contracts, requiring proposal creators to possess a comprehensive understanding of the protocol's functionality and orchestrate the transaction data correctly. This complexity involved in constructing transaction payload discourages community members from actively participating in governance. To tackle this problem, Tally\footnote{https://www.tally.xyz/} provides a user interface for community members to allow them to select from a bunch of supported actions when creating proposals. This approach does encourage communities to engage with the governance process as it is no longer a dilemma to initiate proposals, however, it restricts proposals within limited options and still requires proposal initiators aware of transaction details. Therefore, the highest priority is to reduce the barriers for proposal initiators to generate reliable transaction payload without compromising on flexibility.

\begin{figure}
    \centering
    \includegraphics[width=0.98\columnwidth]{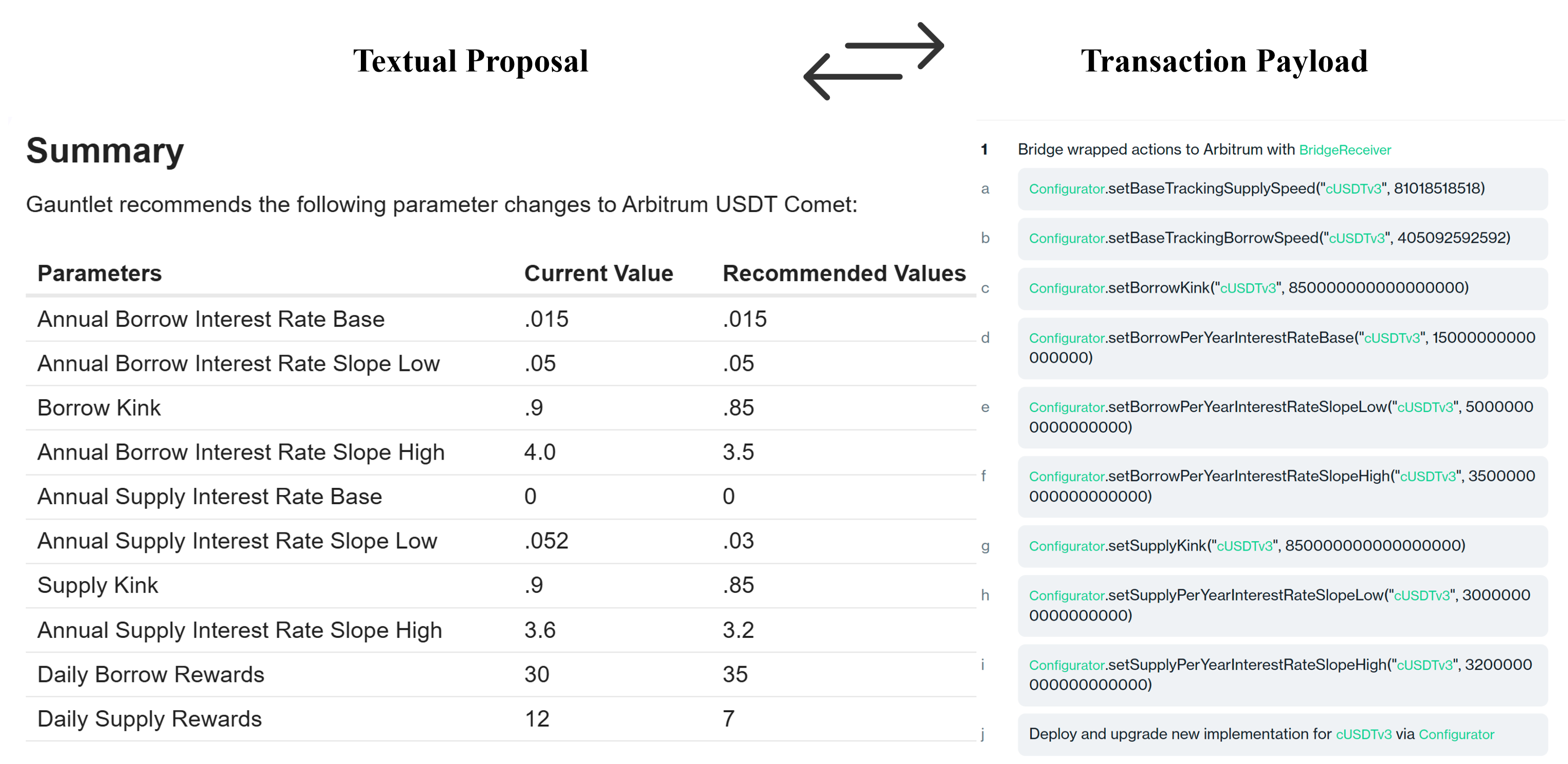}
    \caption{Initate a proposal with the payload on Compound}
    \label{fig:compound}
\end{figure}

\begin{example} \textnormal{Compound is a decentralized lending application relying on collective governance on interest rate models of cryptocurrencies due to market fluctuations. The proposal in figure \ref{fig:compound} suggests updating risk management parameters and incentive setups on the Arbitrium network, with the transaction payload to interact with Arbitrium bridge contract with at least 10 internal message calls. A minor deviation on a single bit would result in financial loss.}
\end{example}

 The \textbf{key insight} in this paper is to propose a novel approach that simplifies the process of initiating transaction payload from textual input. In section \ref{language}, We present the formal definitions of the DAOLang program, specifying its syntax and semantics through Extended Backus-Naur Form (EBNF) and symbolic evaluation. In section \ref{system}, we propose a novel Label-Centric Retrieval (LCR) algorithm and outline the architecture of AgentDAO by detailing how agents collaboratively facilitate program synthesis. The evaluation and experiments are illustrated in section \ref{evaluation}, we reconstruct transaction payload from executed proposals on CompoundV3 and employ an ablation study to analyze the contribution of each component in the system. In section \ref{conclusion}, we review our approach illustrated in this paper, analyzing its merits and limitations.


The \textbf{main contributions} of this paper are summarized as follows.
\begin{itemize}
    \item We propose DAOLang, a high-level and extensive language specialized in expressing proposal actions as intermediate representations for generating proposals.
    \item We present an LLM-based multi-agent architecture to transfer natural languages to proposal payload with zero knowledge of low-level details.
    \item We highlight a novel Label-Centric Retrieval algorithm to semantically realize proposal-sensitive generation.
\end{itemize}

In this paper, we focus on the governance of decentralized lending protocols deployed on multiple blockchains. Our work can be easily migrated to other governance protocols by extending the language. 
\section{The Decentralized Autonomous Governance Language (DAOLang)}
\label{language}

The Decentralized Autonomous Governance Language
(DAOLang) is a high-level commanding language designed
to represent transaction actions in the DAO proposal. It is an LLM-friendly intermediate representation that bridges natural commanding to
transaction data. In this section, we formally define DAOLang by presenting its syntax and symbolic evaluation with illustrative examples. Figure 1 shows a representative set of the syntax of DAOLang.

\subsection{Syntax}
\begin{figure}[h]
\centering
\begin{footnotesize}
    
\begin{align*}
s ::= & & \textbf{Statement:}\\
      & \mid s^* & \text{sequence} \\
      & \mid e; & \text{expression} \\
      & \mid i \leftarrow e; & \text{assignment} \\
      & \mid \nu(\pi : \phi(e^*)); & \text{transaction} \\
e ::= & & \textbf{Expression:}\\
      & \mid i & \text{identifier} \\
      & \mid c & \text{constant} \\
      & \mid o \{i : e, \dots \} & \text{object} \\
      & \mid \nu & network\\
      & \mid \pi & protocol\\
\nu ::= &"network", "\{",<chainId>,"\}"&\textbf{Network.}\\
\pi ::= &"protocol","\{",<address>^*,"\}"&\textbf{Protocol.}\\
\phi ::= &  & \textbf{Actions:} \\
& \mid <addAsset> & addAct\\
& \mid <updateAsset> & updateAct\\
& \mid <operateAsset> & operateAct
\end{align*}
\end{footnotesize}

\caption{A representative set of the syntax of DAOLang} \label{key-components}
\end{figure}
The top level of the DAOLang program consists of a sequence of statements under two categories:
\begin{enumerate}
    \item \textbf{Expressions and Control Flows}: DAOLang encompasses basic and enum expressions. Basic expressions include identifiers, constants, and objects designed for complex data structure, while enum expressions incorporate network and protocol for formatting transactions. Constants and identifiers share consistent type definitions with Solidity\footnote{https://docs.soliditylang.org/en/latest/types.html} types including address, string, signed and unsigned integer, static and dynamic bytes, and boolean. Expressions can be assigned to an identifier through the assignment statement.
    \item \textbf{Transactions and Actions}: Transactions define transaction behaviors by incorporating actions. Inspired by Etherscan\footnote{https://info.etherscan.com/transaction-action/},
    we abstract away miscellaneous information and highlight only key information for complex transactions through the protocol selector  $\pi$ and associated action selector $\phi$. There are three types of actions designed for lending protocols. $addAct$ adds a collateral or a borrowable asset to the protocol.  $updateAct$ changes properties of a collateral or a borrowable asset for management on risk parameters, interest rate models, and incentive setups. $operateAct$ directly operates on a token protocol to move or assign allowance of an asset. The network selector $\nu$ wraps transaction actions with cross-chain logic, indicating on which network the proposal is executed. 
\end{enumerate}

\begin{example} 
    \textnormal{Listing \ref{dsl-example} shows a DAOLang program that proposes updating the supplying cap of wstETH to 5000 in the CompoundV3 WETH Arbitrium market.}
    
\end{example}

\begin{lstlisting}[caption={DAOLang example for updating the supply cap of collateral in a proposal.},label={dsl-example}]
# assign the address of wstETH to a variable
collateral <- 0x7f39c581f595b53c5cb19bd0b3f8da6c935e2ca0
# assign the address of WETH to a variable
market <- 0x82aF49447D8a07e3bd95BD0d56f35241523fBab1;
# declare a network and assign it to a variable
arb <- network{42161}
# declare a protocol and assign it to a variable
compV3 <- protocol{0x316f9708bB98af7dA9c68C1C3b5e79039cD336E3}
# Update the supplying cap of wstETH
arb(compV3:update_supplyCap(collateral,market,5000));
\end{lstlisting}

\subsection{Symbolic Evaluation}

\begin{figure}
\begin{footnotesize}
    
\[
\frac{
\forall i \in \{1,\dots,n\}, \quad
\langle s_i, \delta_{i-1}, \tau_{i-1} \rangle \xrightarrow{} \langle \emptyset, \delta_i, \tau_i \rangle
}{
\langle (s_0, \dots, s_n),  \delta, \tau \rangle \xrightarrow{}  \langle \emptyset, \delta_n, \tau_n \rangle
}
\quad \text{\scriptsize (SEQN)}
\]

\[
\begin{aligned}
&\frac{
\phantom{}
}{
\langle c, \delta, \tau \rangle \xrightarrow{} \langle c, \delta, \tau \rangle
}
\quad \text{\scriptsize (CNST)}
&&\frac{
\phantom{}
}{
\langle i, \delta, \tau \rangle \xrightarrow{} \langle \delta[i], \delta, \tau \rangle
}
\quad \text{\scriptsize (IDEN)}
\end{aligned}
\]
\[
\frac{
\renewcommand{\arraystretch}{1.2}
\begin{array}{c}
\forall i \in \{1, \dots, n\}, \quad
\langle e_i, \delta, \tau \rangle \xrightarrow{} \langle v_i, \delta, \tau \rangle 
\end{array}
}{
\langle o, \delta, \tau \rangle \xrightarrow{} \langle \{k_1:v_1, \dots, k_n:v_n\}, \delta, \tau \rangle
}
\quad \text{\scriptsize (OBJT)}
\]

\[
\frac{
\begin{aligned}
&\langle e, \delta, \tau \rangle \xrightarrow{} \langle v, \delta_0, \tau \rangle \\
&\delta' = \delta_0 \cup \{i \mapsto  v\}
\end{aligned}
}{
\begin{aligned}
\langle i \leftarrow e, \delta, \tau \rangle \xrightarrow{} \langle \emptyset, \delta', \tau \rangle
\end{aligned}
}
\quad \text{\scriptsize(ASGN)}
\]
\[
\frac{
\renewcommand{\arraystretch}{1.2}
\begin{array}{c}
\forall i \in \{1, \dots, n\}, \quad
\langle \nu(t_i), \delta, \tau \rangle \xrightarrow{} \langle \emptyset, \delta, \tau_i \rangle \\
\langle \pi:\phi(e^*), \delta, \tau \rangle \xrightarrow{} \langle \emptyset, \delta, \tau_i \rangle\\
 \tau' = \mathcal{N}(\bigcup_i \tau_i) \quad if \: \nu \in \mathcal{V} \quad else \: \bigcup_i \tau_i
\end{array}
}{
\langle \nu(t_0,\dots,t_n), \delta, \tau \rangle \xrightarrow{} \langle \emptyset, \delta, \tau' \rangle
}
\quad \text{\scriptsize(TRAN)}
\]

\[
\frac{
\renewcommand{\arraystretch}{1.2}
\begin{array}{c}
\langle \pi:\phi(e^*), \delta, \tau \rangle \xrightarrow{} \langle \pi:\mathcal{E}(\phi(e^*)), \delta, \tau \rangle \\
\langle \pi, \delta, \tau \rangle \xrightarrow{} \langle \emptyset, \delta, \tau^\pi \rangle \quad if \: \pi \in \Pi \quad else \: Err  \\
\langle \mathcal{E}(\phi(e^*)), \delta, \tau \rangle \xrightarrow{} \langle \emptyset, \delta, \tau^\phi \rangle \quad if \: \phi \in \Phi(\pi) \quad else \: Err \\
\tau' = \tau^\pi \cup \tau^\phi
\end{array}
}{
\langle \pi:\phi(e^*), \delta, \tau \rangle \xrightarrow{} \langle \emptyset, \delta, \tau' \rangle
}
\quad \text{\scriptsize(ACTN)}
\]
    \caption{A representative set of the symbolic evaluation of the DAOLang program}
    \label{fig:semantics}
\end{footnotesize}  
\end{figure}

We describe how the DAOLang program is symbolically evaluated via a set of evaluation rules. The state of a DAOLang program can be represented by a 3-tuple $\langle p,\delta,\tau\rangle$, where:
\begin{enumerate}
    \item $p$ is the program counter that points to the next DAOLang statement.
    \item $\delta$ is the program store that provides access to program memory.
    \item $\tau$ encompasses a set of bytecode to form a transaction object. Specifically, we design a transaction payload as the program output to indicate how proposals should interact with smart contracts.
\end{enumerate}

Figure \ref{fig:semantics} describes a representative set of symbolic evaluation rules. $\langle p,\delta,\tau \rangle \xrightarrow{} \langle p', \delta', \tau'  \rangle$ denotes a transition from program state $\langle p,\delta,\tau \rangle$ to program state $\langle p', \delta', \tau'  \rangle$. The evaluation process starts with (SEQN) rule, which populates and evaluates statement $s$ sequentially. We denote by (CNST), (IDEN), and (OBJT) three different rules to retrieve data via offering constant value, accessing the program store $\delta$, and combing (CNST) and (IDEN) for structural sets. The assignment rule (ASGN) binds an identifier $i$ to a location in the program store $\delta$.

The (TRAN) and (ACTN) rules illustrate the process of generating the transaction payload during the execution of the DAOLang program. $\pi$ and the function $\mathcal{E}(\phi(e*))$ initiate a set of transactions and $\mathcal{N}(\bigcup_i \tau_i)$ wrap these transactions with cross-chain logic. Specifically, $\pi$ maps the protocol name to contract addresses, and $\mathcal{E}(\phi(e*))$ converts actions to their corresponding function signatures and encodes parameters into bytecode based on their types. The (ACTN) rule ensures that the protocol addresses are within a valid set defined in $\Pi$ and the function signatures generated by $\mathcal{E}(\phi(e*))$ are within defined in $\Phi(\pi)$. After verifying that $\nu$ is a member of the valid networks $\mathcal{V}$, $\mathcal{E}(\bigcup_i \tau_i)$ integrates the outputs of (ACTN), resolving them to a cross-chain transaction.



\begin{example}
\textnormal{Listing \ref{dsl-example-2} shows how (TRAN) and (ACTN) rules generate the transaction payload for a given DAOLang program input. First, the (ACTN) rule resolves the program into two sequential transactions that interact with the Configurator (0x316f) and CometProxyAdmin (0x1EC6) contracts respectively, with embedding the contract addresses into the transaction payload and specifying their function signatures as updateAssetSupplyCap and deployAndUpgradeTo. Subsequently, (ACTN) encodes the parameters from $\delta$ based on their respective types and populates the payload in the payload. The (TRAN) rule encapsulates the above transactions into the payload of a bridge message call, which interacts with the Inbox (0x4Dbd) contract to bridge to the Arbitrum network by invoking the createRetryableTicket function.
}
\end{example}


\begin{lstlisting}[caption={Example for generating transaction payload.}, label={dsl-example-2}]
|\textbf{Program input:}|
# Update the supplying cap of RsETH
|\textcolor{red}{arb}|(|\textcolor{orange}{compV3}|:|\textcolor{blue}{update\_supplyCap(collateral,market,5000)}|);
|\textbf{Transaction payload:}|
[{
|\textcolor{red}{address}|:0x4Dbd4fc535Ac27206064B68FfCf827b0A60BAB3f}
value:0
|\textcolor{red}{functionSig}|: createRetryableTicket(address,uint256,
uint256,address,address,uint256,uint256,bytes)
|\textcolor{red}{payload}|: [
    {
    |\textcolor{orange}{address}|:0x316f9708bB98af7dA9c68C1C3b5e79039cD336E3
    value:0
    |\textcolor{blue}{functionSig}|:updateAssetSupplyCap(address,address,uint128)
    |\textcolor{blue}{payload}|:0x000000000000000000000000f176fb51f4eb826136a54f
    dc71c50fcd2202e2720000000000000000000000000fbcbaea96ce0c
    f7ee00a8c19c3ab6f5dc8e1921000000000000000000000000000000
    0000000000000000000000000000001388
    },{
    |\textcolor{orange}{address}|:0x1EC63B5883C3481134FD50D5DAebc83Ecd2E8779
    value:0
    |\textcolor{blue}{functionSig}|:deployAndUpgradeTo(address,address)
    |\textcolor{blue}{payload}|:0x000000000000000000000000316f9708bb98af7da9c68c
    1c3b5e79039cd336e30000000000000000000000000fbcbaea96ce0c
    f7ee00a8c19c3ab6f5dc8e1921
    },
    ...
]

\end{lstlisting}

\section{System Design}
\label{system}
\begin{figure*}[h!]
    \centering
    \includegraphics[width=\textwidth]{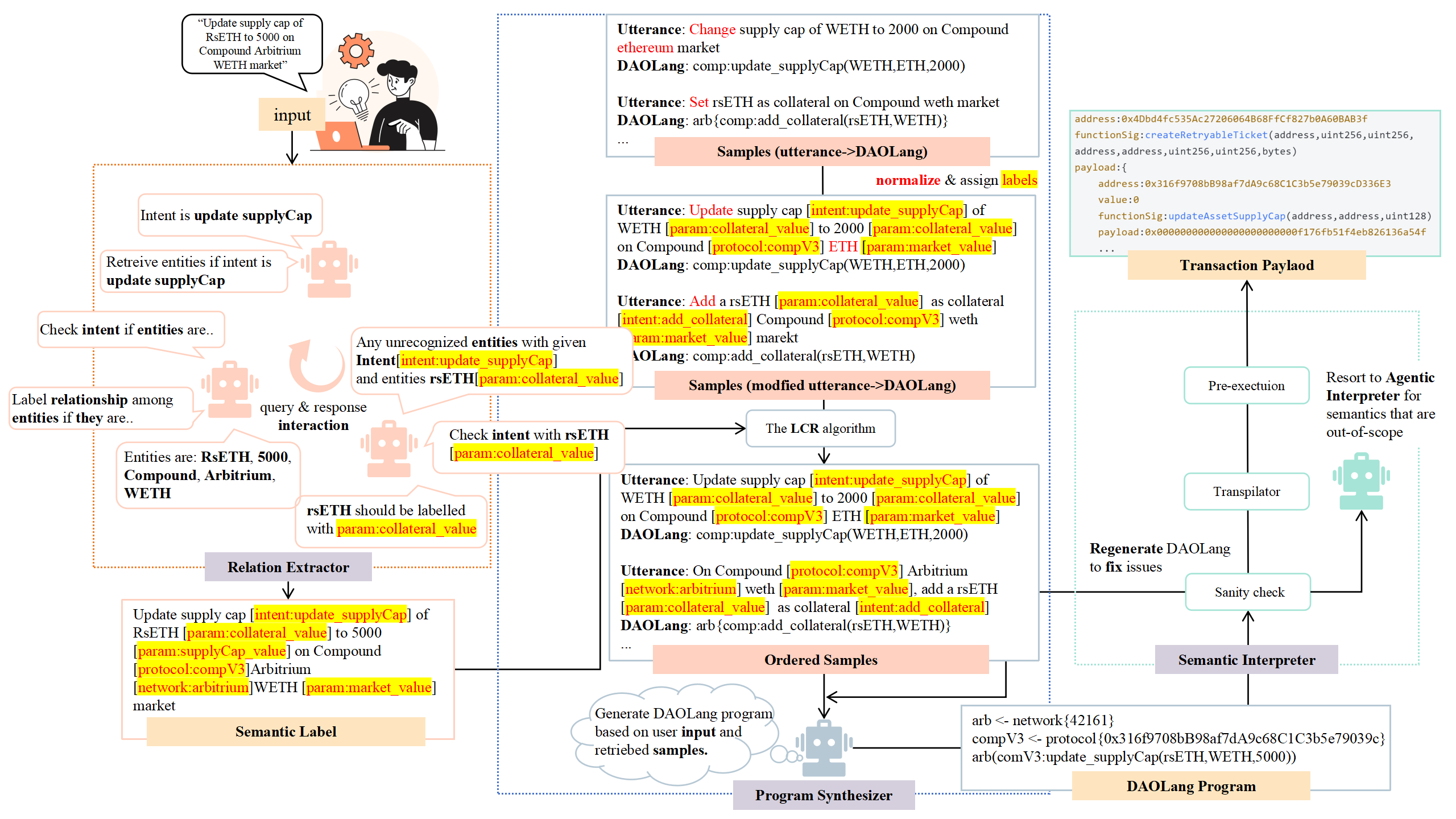}
    \caption{Illustration of the architecture of AgentDAO. User utterance is translated into the DAOLang program and is reformulated into a transaction payload.}
    \label{fig:system}
\end{figure*}
We illustrate the design and evaluation process of DAOLang in section \ref{language}. In this section, we present our approach using
LLMs to synthesize DAOLang programs from textual inputs. 
Figure \ref{fig:system} illustrates the overall architecture of AgentDAO, which incorporates LLM-based agents under three fundamental layers: Relation Extractor (RE), Program Synthesizer (PS), and Semantic Interpreter (SI). These agents have distinct roles and communicate with each other in a hierarchical structure. RE takes natural language as input and expands user utterance with semantic role labels. PS selects samples via a novel Label-Centric Retrieval (LCR) algorithm to synthesize the DAOLang program. After validating the program, SI transpiles it into transaction payloads and simulates their execution to ensure the transaction behavior aligns with expectations.

\subsection{Relation Extractor}
User utterance is expanded and enriched with three agentic components: intent classifier, entity extractor, and semantic role labeler. The intent classifier is responsible for identifying the overarching objective or intended action behind the user’s query, such as interacting with a token or managing risk parameters. The entity extractor systematically extracts essential components within the input, such as parameters, attributes, or contextual elements. Building on these, the labeler assigns extracted entities appropriate semantic roles to represent their relationships. We reformulate the relation extraction task into a query-response framework based on SUMASK \cite{li-etal-2023-revisiting-large} and adapt the approach to the multi-agent system to enhance agent collaboration by structuring interactions. This allows agents to sequentially share and refine information, ensuring that each agent's output informs and builds upon the others, fostering efficient and coordinated task completion within the multi-agent system.

\subsection{Program Synthesizer}
In this section, we describe our approach to synthesizing the DAOLang program via prompt construction. Given modified utterances with semantic labeling from RE, we introduce a Label-Centric Retrieval (LCR) method to improve the performance of the semantic search process to choose similar sample utterances and the corresponding DAOLang programs. The process consists of six steps:

\paragraph{Samples Construction.} We construct a sample database $\mathcal{D}$ incorporating pairs of DAOLang program samples with corresponding utterances. Each program is annotated with function descriptions and parameter specifications. 

\paragraph{Normalization.} To prevent redundant matches, sample utterances are normalized by standardizing all intent-descriptive actions and smart contracts-related entities. For example, we use "update" to replace other descriptive actions including "change", "configure", and "set", and use symbols of token contracts to replace all alternative representations. 

\paragraph{Assigning Labels.} To facilitate the retrieval algorithm described in \ref{alg:algorithm}, we assign labels to entities in utterance samples to highlight the intended action and indicate relationships among intents, function signatures, function parameters, networks, and associated protocols. Labels are also normalized.

\paragraph{Embedding.} Normalized utterance $\hat{u}_i$
and user utterance $x$ are embedded to vectors $\mathcal{E}(\hat{u}_i)$ and $\mathcal{E}(x)$ based on dense passage retrieval (DPR) \cite{karpukhin-etal-2020-dense}.  The core idea is to map both the query and the payload passages into a high-dimensional vector space where semantic similarities can be measured using cosine similarity. In our work, the distance between $\hat{u}_i$ and $x$ in the vector space is denoted by $\|\mathcal{E}(\hat{u}_i)-\mathcal{E}(x)\|$.


\paragraph{The LCR algorithm.} The key insight behind Label-Centric Retrieval (LCR) is to “teach” an LLM the most relevant language spec w.r.t. a given input. This is achieved via retrieving an ordered list of samples which match the highest level of semantics from the input labels, e.g., require an exactly same function call. The LCR algorithm is illustrated in algorithm \ref{alg:algorithm}. We retrieve sample set $\mathcal{S}$ with up to $k$ samples from the sample database $\mathcal{D}$, ranked in descending order according to the marginal contribution to the accrued number of matched labels. In the case of equal marginal contributions, the samples are ranked according to their distances from $\mathcal{E}(x)$ in the vector space. The retrieved set $\mathcal{S}$ always collects the largest number of matched labels with the smallest $k$ and maximizes the potential number of matched labels for any given $k$. Clearly, samples with duplicate or subset labels are given a lower priority, even if they are close to user input in the vector space.

\paragraph{Prompt Construction. } We integrate the system prompt with retrieved samples for LLMs to generate the DAOLang program. The system prompt encompasses two parts: specifications of DAOLang and step-by-step instructions for program synthesis.

\begin{figure}
    \centering
    \includegraphics[width=0.98\columnwidth]{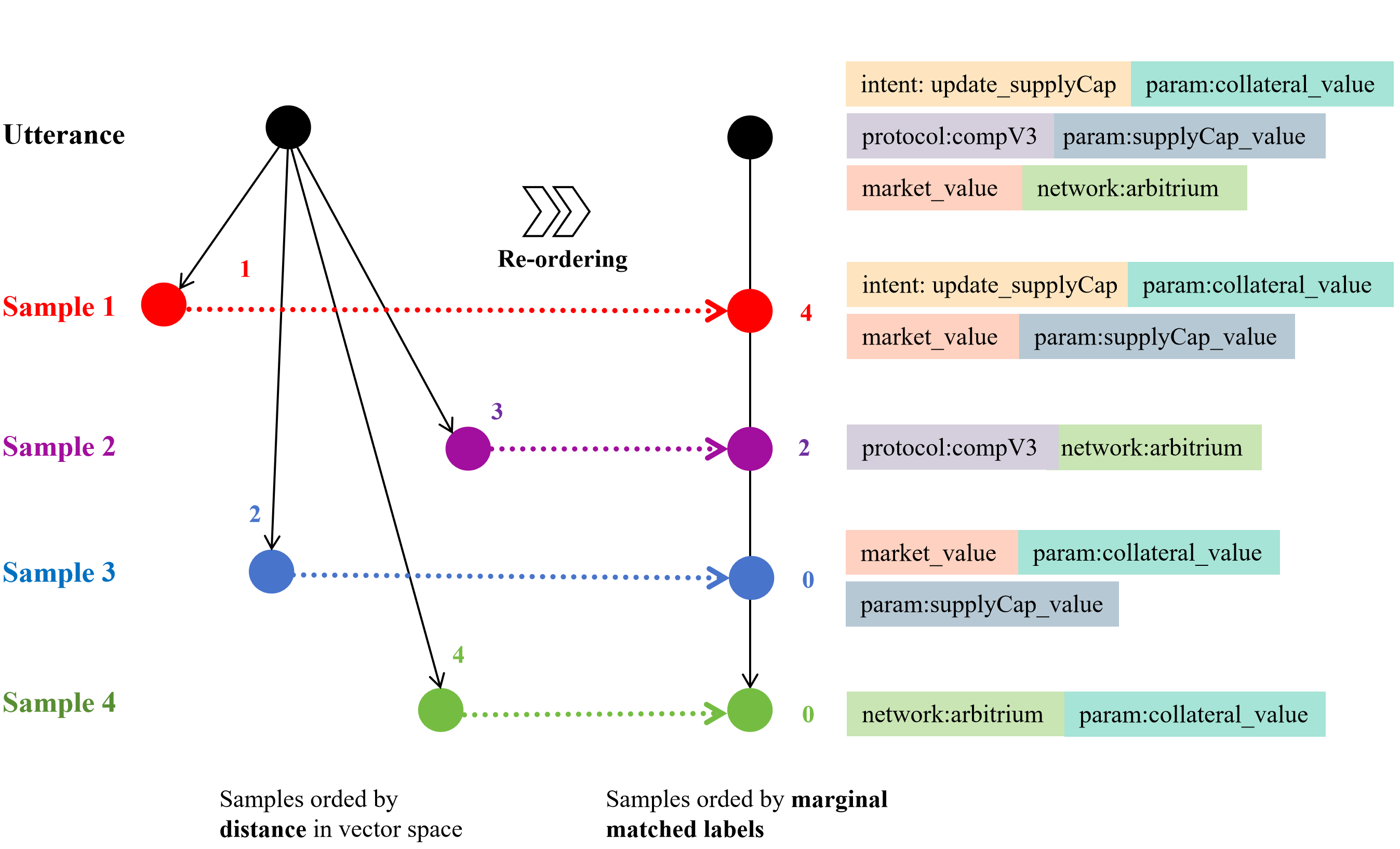}
    \caption{An illustrative example of LCR. The LCR algorithm first ranks samples according to their vector space and then ranks them  in descending order based on the marginal contribution matched labels}
    \label{fig:LCR}
\end{figure}

\begin{example}
\label{LCR-example}
\textnormal{Figure \ref{fig:LCR} describes the optimal ranking order of 4 utterance samples computed by the LCR algorithm. Any other combination has fewer accrued matched labels when $k=1$ or $k=2$. Sample 3 and Sample 4 are ordered according to their distance to utterance in the vector space as they have equal marginal contribution on matched labels.  }
\end{example}

\begin{algorithm}
\caption{The Label-Centric Retrieval algorithm.}
\label{alg:algorithm}
\begin{algorithmic}[1]
    \REQUIRE User input \(x\), sample database \(\mathcal{D} \), and maximum number of examples to retrieve \(k\)
    \ENSURE Retrieved samples \( \mathcal{S} \)
    \STATE List of labels in user input \( \hat{\mathcal{L}} \xleftarrow{} \mathcal{L}(x) \)
    \STATE List of relevant samples \( \mathcal{R} \xleftarrow{} \{\} \)
    \STATE Mapping from utterance to distance \( \mathcal{M} \xleftarrow{} \{\} \)
    \FOR{\(\hat{u}_i \in \mathcal{D}\)}
        \IF{\(\mathcal{L}(\hat{u}_i) \cap \hat{\mathcal{L}} \neq \emptyset\)}
            \STATE \(\mathcal{R}.insert(\hat{u}_i)\)
        \ENDIF
    \ENDFOR
    \FOR{\(\hat{u}_i \in \mathcal{R}\)}
        \STATE \(\mathcal{M}(\hat{u}_i) \xleftarrow{} \|\mathcal{E}(\hat{u}_i)-\mathcal{E}(x)\|\)
    \ENDFOR
    \STATE \( \mathcal{R} \xleftarrow{} \mathcal{R}.sortBy(\mathcal{M}(\hat{u}_i), asc)  \)
    \FOR{\(\mathcal{R} \neq \emptyset\)}
        \STATE \(j=0,max=0\)
        \FOR{\(\hat{u}_i \in \mathcal{R}\)}
            \STATE curr = \((\mathcal{S} \cup \mathcal{L}(\hat{u}_i)).length  \)
            \IF{\(curr > max\)}
                \STATE \(max = curr\)
                \STATE \(j=i\)
            \ENDIF
        \ENDFOR
        \STATE \( \mathcal{S}.append(\hat{u}_j)\)
        \STATE \( \mathcal{R}.remove(\hat{u}_j)\)
    \ENDFOR
    \STATE \( \mathcal{S} \xleftarrow{} \mathcal{S}[0:max(k,\mathcal{S}.length)]  \)
    \STATE \textbf{Return} \( \mathcal{S} \)
\end{algorithmic}
\end{algorithm}

\subsection{Semantic Interpreter}
SI validates the DAOLang program and transpiles it to the transaction payload. The validation process consists of two stages, each focusing on DAOLang programs and generated transaction payloads. The sanity check parses the program into an Abstract Syntax Tree (AST) to ensure syntactic correctness and semantic consistency. The second stage simulates blockchain states within a local environment to verify that the execution outcomes of the generated proposal transactions are in line with the original utterance. If any error happens, SI recalls PS with error information to regenerate the DAOLang program to fix issues.

In section \ref{language}, we have discussed how DAOLang programs are transpiled to transaction payloads by illustrating the symbolic evaluation process, which requires protocols $\pi$ and actions $\phi$ are within finite sets $\Pi$ and $\Phi$. However, there are occasional cases where proposal actions are atypical and not well-defined in DAOLang. In this case, we present a tool-learning-based agentic interpreter that retrieves ABI files from the Etherscan API. The interpreter then matches function signatures in the ABI files to generate transaction payloads, taking into account both the synthesized DAOLang programs and the original user input.

\section{Evaluation}
\label{evaluation}
In Section \ref{system}, we described our system as incorporating LLM-based agents across three fundamental layers, each providing distinct functionalities to the overall system. To comprehensively evaluate the performance of AgentDAO and analyze the contributions of agents within each layer, we evaluate the system at variant $k$ (The maximum number of retrieved samples) and conduct an ablation study by removing agents from the system layer by layer. Our system utilizes the GPT-4o model, supporting prompt constructions of up to 128,000 tokens. 

\paragraph{Measurement.}We select 155 executed DAO proposals from CompoundV3 Governance protocol between 2023 and 2024 and reconstruct transaction payloads from their transaction data as a benchmark to evaluate transaction payloads generated by agentDAO. The original evaluation procedure is then reformulated to equivalence analysis. We give four match levels as follows:
\begin{enumerate}
    \item \textbf{Exact Match (EM)}: The transaction payload of the generated proposal is identical to the standard transaction payload.
    \item \textbf{Functional Equivalence (FE)}: The transaction payload of the generated proposal is inconsistent with the standard transaction payload, but demonstrates identical state changes on the blockchain.
    \item \textbf{Semantic Analogy (SA)}: The generated proposal's transaction payload deviates from the standard, but its execution results in semantically comparable state changes. We denote SA as a failure case, indicating that the generated proposal does not meet the required standards.
    \item \textbf{Error Identification (EI)}: The generated proposal's transaction payload is inconsistent with the standard, or AgentDAO fails to produce a valid transaction payload.
\end{enumerate}

\begin{table*}[h!]
\caption{Results of experimental studies, running on the evaluation procedure discussed in section \ref{evaluation}.}
\label{table:results}
\centering
\small 

\renewcommand{\arraystretch}{1.15} 
\setlength{\tabcolsep}{6pt} 
\begin{tabular}{ccccccccr}
\hline
\textbf{$k$}&\textbf{RE}&\textbf{PS}&\textbf{SI}&  
\textbf{Exact Match} & \textbf{Functional Equivalence} & \textbf{Semantic Analogy} & \textbf{Error Identification} & \textbf{Pass Rate (\%)} \\ \hline

0&$\checkmark$&$\checkmark$&$\checkmark$  & 8  & 3  & 0 & 144 & $7.01$  \\ 
2&$\checkmark$&$\checkmark$&$\checkmark$  & 72 & 22  & 3 & 58  & $60.65$  \\ \hline
4&$\checkmark$&$\checkmark$&$\checkmark$  & 102 & 37  & 4 & 12  & \textbf{89.68}  \\ \hline
5&$\checkmark$&$\checkmark$&$\checkmark$  & 97 & 39  & 4 & 15  & $87.74$  \\ 
6&$\checkmark$&$\checkmark$&$\checkmark$  & 96 & 38  & 4 & 17  & $86.45$  \\ 
2&$\times$&$\checkmark$&$\checkmark$ & 53 & 19  & 4 & 44 & $46.45$  \\ 
4&$\times$&$\checkmark$&$\checkmark$ & 82 & 34  & 4 & 35 & $74.84$  \\ 
5&$\times$&$\checkmark$&$\checkmark$ & 91 & 32  & 4 & 28 & \textbf{79.35}  \\ 
6&$\times$&$\checkmark$&$\checkmark$ & 89 & 32  & 4 & 30 & $78.06$  \\ 
-&$\checkmark$&$\times$&$\checkmark$ & 16 & 3  & 3 & 133 & \textbf{12.26}  \\ 
2&$\checkmark$&$\checkmark$&$\times$ & 65 & 18  & 3 & 50 & 53.55  \\ 
4&$\checkmark$&$\checkmark$&$\times$ & 95 & 33  & 4 & 19 & \textbf{82.58}  \\ 
5&$\checkmark$&$\checkmark$&$\times$ & 90 & 35  & 4 & 22 & $80.64$  \\ \hline
\end{tabular}
\end{table*}

\subsection{Preliminary Evaluation.} Table \ref{table:results} shows that AgentDAO, with all components enabled, achieves its highest pass rate of $89.68\%$ at $k=4$. The zero-shot prompting without any DAOLang samples results in a low pass rate of around 7\% while providing too many samples (setting k to a larger number) brings slight negative impacts due to increasing noise and overfitting to retrieved data. 
\paragraph{Merits of agents in each layer.} 
 When RE is removed, the system achieves its highest pass rate of $79.35\%$ at $k=5$. As a result of removing RE, the LCR algorithm regresses to the standard retrieval method due to the absence of semantic role labels. When either PS or the agentic interpreter in SI is removed from the system, the highest pass rate drops to $12.26\%$ and $82.58\%$, respectively. The result is in line with our expectation, as DAOLang is designed to express canonical proposal actions, while the agentic interpreter is used to handle uncommon actions that are inexpressible through DAOLang.


\paragraph{Inefficiency in gas optimization}
AgentDAO generates numerous proposals that achieve functional equivalence to standard proposals. However, while these synthesized proposals produce identical state changes on the blockchain, they often lack conciseness in terms of gas optimization. For example, proposal\footnote{https://compound.finance/governance/proposals/161} aims to set supply caps for collaterals and update the market incentive rate within a single bridge message. In contrast, the proposal synthesized by AgentDAO distributes these actions across multiple bridge messages. Although both approaches ultimately execute as atomic transactions, the latter incurs significantly higher gas costs due to the increased number of bridge operations.

\paragraph{Challenges in Transforming Ambiguous or Undefined Commands.}
The experimental results indicate that AgentDAO fails to translate 16 out of 155 textual proposals into the expected transactions. These failures primarily stem from proposals with ambiguous command descriptions or those involving actions that cannot be expressed via DAOLang.

\subsection{Categorizing Proposals} 
\begin{figure}
    \centering
    \includegraphics[width=0.98\columnwidth]{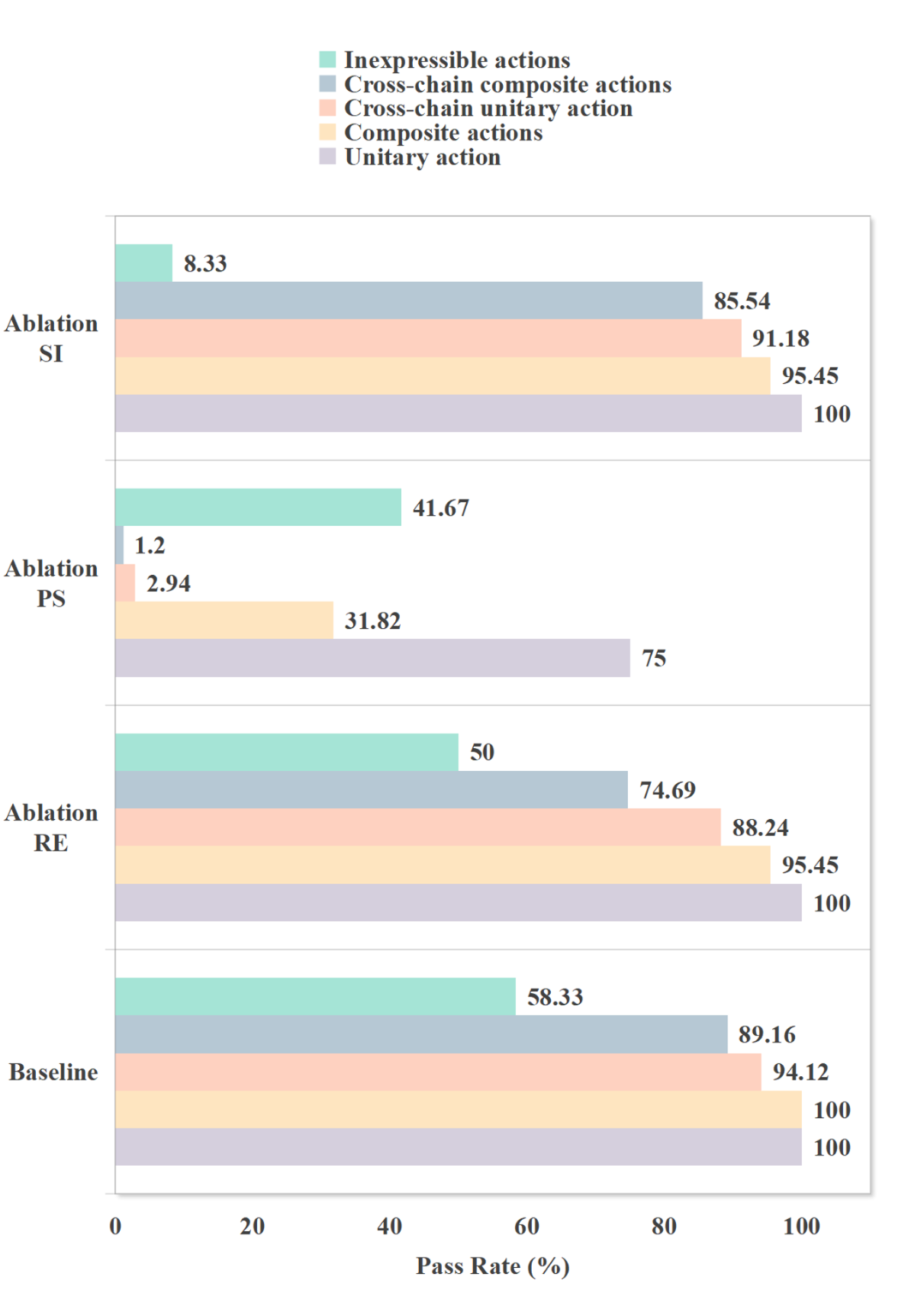}
    \caption{An ablation study on system performance across categories.}
    \label{fig:ablation}
\end{figure}

We further categorize the 155 selected CompoundV3 proposals into five distinct categories based on the types of action of the proposal. The collected dataset consists of 4 proposals with unitary actions, 22 with composite actions, 34 with cross-chain unitary actions, 83 with cross-chain composite actions, and 12 with inexpressible actions. AgentDAO, with all agents enabled, is used as a baseline to evaluate the performance of each ablation within these categories. The results of the ablation study across these categories are presented in Figure \ref{fig:ablation}.

\paragraph{Advantages in expressing composite 
 and cross-chain instructions.} Ablating the PS component significantly degrades the system's performance, especially for proposals involving composite actions and cross-chain requests. Specifically, the pass rates for synthesizing proposals drop from 87.96\% to 1.2\% for cross-chain composite actions, from 94.12\% to 2.94\% for cross-chain unitary actions, and from 100\% to 31.82\% for composite actions. These results underscore the critical role of integrating the DAOLang program for expressing proposal actions, highlighting its fundamental importance to the system’s overall effectiveness.

 \paragraph{Synergies of RE and the LCR algorithm.} The absence of RE forces the PS component to handle user utterances without labeling entity relationships, causing the LCR algorithm to revert to standard distance-based retrieval methods due to the lack of semantic role labels. Ablating RE influences proposal synthesis across all categories, except for unitary actions, as described in figure \ref{fig:ablation}. Table \ref{table:results} shows that the system without RE achieves its best performance with a larger $k$ compared to the baseline, suggesting the synergy of RE and the LCR algorithm outperforms DPR in selection quality and token efficiency under the context of program synthesis.

 \paragraph{Limitations and out-of-scope issues.} The pass rate for synthesizing proposals involving inexpressible actions declines from 58.3\% to 8.33\% when the agentic interpreter is removed from the SI system, resulting in an inability to handle actions that are out of the scope of the DAOLang program. While extending DAOLang is a viable option, the tool-learning-based interpreter offers some flexibility in understanding simple and arbitrary commands.

\section{Related Work}
\paragraph{LLM-based program synthesis.}
Recent advances in large language models (LLMs) have enabled using natural language as a specification for the automatic generation of programs. While CodeGen \cite{nijkamp2022codegen}, PaLM \cite{chowdhery2023palm}, and Codex \cite{chen2021evaluating} are LLM-based tools that generate general-purpose programming languages from natural language description, other work \cite{gandhi2023natural} leveraged LLMs to generate a domain-specific language for operating Office applications. LLMs demonstrate promising capabilities in program synthesis, with performance scaling predictably with size and fine-tuning, but challenges remain in achieving deeper semantic understanding and consistent reliability \cite{DBLP:journals/corr/abs-2108-07732}. In this paper, we employ a multi-agent system to synthesize DaoLang programs, an expressive and succinct domain-specific language, from natural language specifications. To ensure the reliability and accuracy of the generated code, we further integrate static validation techniques and dynamic simulation processes into the synthesis pipeline.
\paragraph{ICL-based relation extraction.}
Relation extraction (RE) is a critical task that involves identifying relationships between entities in text. While large language models (LLMs) have showcased remarkable in-context learning (ICL) capabilities \cite{NEURIPS2020_1457c0d6}, recent research reveals persistent challenges in zero-shot and few-shot RE scenarios. To address these issues, researchers have explored various strategies, such as designing effective query forms \cite{li-etal-2023-revisiting-large} and employing retrieval-based techniques \cite{DBLP:journals/corr/abs-2303-08559} to improve ICL-based RE performance. Recent advancements include a meta-training framework \cite{ijcai2024p702} that fine-tunes LLMs for ICL using diverse RE datasets, a recall-retrieve-reason framework \cite{ijcai2024p704} that combines LLMs with retrieval corpora to enhance relevance and reasoning, and a novel attention-based method \cite{ijcai2024p726} for document-level RE. In our work, RE serves as a foundational component of the DAOLang program synthesis framework. We leverage a similar approach with the SUMASK (Summarize-and-Ask) prompting framework \cite{li-etal-2023-revisiting-large} and adapt the framework to the multi-agent system to enable the query and response interaction. 


\section{Conclusion}
\label{conclusion}


In this paper, we present a multi-agent system that uses natural languages as input to produce transaction payload. The system incorporates numerous agents with distinct roles in a hierarchical structure to transfer utterances to the DAOLang program and subsequently transpile the program to transaction payloads. The program synthesis process employs a Label-Centric Retrieval (LCR) algorithm, collaborating with the Relation Extractor to facilitate prompt construction. Experiments suggest that the collaboration of LLM-based agents in AgentDAO and the introduction of DAOLang significantly improve the system's capability, especially in orchestrating transactions involving composite commands and cross-chain interactions, even though generated transactions are not always concise enough considering gas consumption. 

Our work demonstrates the potential of organizing multiple LLM-based agents and offers insights into simplifying complex tasks for users with zero-knowledge of low-level details.

\newpage

\bibliographystyle{named}
\bibliography{ijcai25}
\end{document}